\documentstyle[aps,twocolumn,graphics]{revtex}
\newcommand{\ra}{\rangle}
\newcommand{\la}{\langle}
\newcommand{\om}{\omega}

\begin{document}
\draft

\wideabs{
\title{\bf Collective Rabi Oscillations and Cold Collisions.}
\author{J.I. Kim and P. Nussenzveig}
\address{Instituto de F\'\i sica, Universidade de S\~ao Paulo, {\protect \\}
Caixa Postal 66318, CEP 05315-970, S\~ao Paulo, SP, Brazil.}
\date{June 2001}
\maketitle
\begin{abstract}
We examine cold atomic collisions within a resonant optical cavity. The
quantized cavity mode can be used to manipulate the collisions between the cold 
atoms, such that periodic exchange of excitations between the atoms and the
electromagnetic field strongly alters the collision dynamics. A colliding pair 
of atoms can thereby oscillate between its ground and excited states during the 
collision time. Using a semiclassical model, it can be predicted that such
Rabi-like oscillations are revealed in the atomic trap-loss probabilities,
which show maxima and minima as a function of the detuning between the
frequencies of the mode and the atomic transition.
\end{abstract}
}

\noindent
Exchange of excitations between the energy levels of atoms or
molecules and the quantum electromagnetic field \cite{yoo85}
is one basic interaction process between matter and light. In this
context, the Jaynes-Cummings model \cite{jc63}, describing a single two-level
system and a monochromatic lossless field, reveals several characteristics of
this interaction. In the quantum Rabi
oscillation \cite{haroche96,wine96}, in particular, a single energy quantum
is periodically exchanged between system and field. On the other hand, when
several atoms or molecules interact with the same field, quantum coherence
can build among them leading to well known collective behaviors, such as
superradiance \cite{dicke54,bonifacio71,gibbs81,gross82,haroche82}, for which
multiparticle entanglement plays a major role. It should be noticed, however,
that the systems with which the field interacts need not be solely composed of
stable atoms or molecules. The above characteristics of the matter-field
interaction can be present as well when two cold atoms collide in presence of a
quantum field. An additional effect is that the very dynamics of the cold
atomic collision \cite{gp89,weiner99} can be strongly modified by this field.
Indeed, pairs of colliding atoms so far apart from each other that their
direct mutual interaction is negligible, can be entangled by the field and,
thereby, influence one another in a nonlocal way. In the following we study 
cold atomic collisions within a gas
of cold atoms trapped in the center of a high-$Q$ cavity. It is possible to
show that an analogous {\em collective Rabi oscillation} can show up in the
trap-loss probabilities as a function of the detuning between the cavity mode
frequency and the atomic resonance.

\begin{figure}
\centering \resizebox{7.5cm}{!}{\includegraphics*{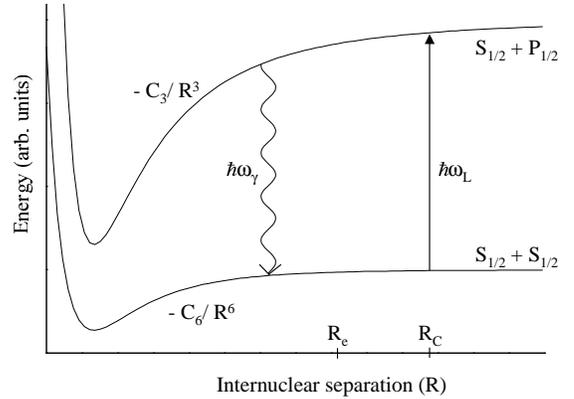}}
\caption{Excited state long range dipole-dipole attractive 
potential $U=-C_3/R^3$ and the ground state van der Waals 
potential $1/R^6$, as functions of the internuclear distance
$R$. The energy difference of the latter to the asymptote of $U$ is
the atomic separation $\hbar\om_A$ between $nS_{1/2}$ and $nP_{1/2}$.}
\label{fig:potential}
\end{figure}

The internuclear potential energy of two colliding atoms depends on
the electronic states. In particular, the inverse-cube law
dipole-dipole potential $\pm 1/R^3$ between alkali neutral atoms,
separated by a distance $R$, predominates when the asymptotic atomic
states involved are $nS_{1/2}$ and $nP_{1/2}$
(Fig.~\ref{fig:potential}). When the collision is slow
\cite{gp89,weiner99}, the atoms can undergo changes in their electronic 
states during the collision, either as spontaneous decay or induced
transitions. In a previous work \cite{lowq00}, it was examined how
cold collisions can be manipulated as spontaneous decay is driven by
the colored vacuum of a cavity. The highly increased emission rate of
multiply entangled pairs being able to emit coherently to the same
cavity mode was then predicted to practically interrupt the collision
process. In the present case, we neglect the cavity loss (high-$Q$) and
allow reabsortions of the field energy. We study how these induced 
transitions affect cold collisions.  
Cavity Quantum Electrodynamics (CQED) effects on atomic motion in
high-$Q$ cavities have been investigated recently. Modifications of
mechanical forces of light acting on atoms \cite{mossberg91,gangl00},
or appearance of new quasi-bound molecular states of two colliding
atoms \cite{deb99} illustrate the interplay between CQED and cold
atoms.

A pair of colliding atoms under the attractive potential
$U(R)=-C_3/R^3$, with $C_3$ a constant, is described as in \cite{lowq00} by a
two-level system. Its energy splitting $\hbar\om_R=\hbar\om_A+U(R)$ depends on 
the internuclear distance $R$ and approaches the atomic energy difference
$\hbar\om_A$ between $nS_{1/2}$ and $nP_{1/2}$ as 
\mbox{$R\longrightarrow\infty$}. In this asymptotic limit, the excited
state of the pair denoted by $|e\ra$ correlates to $nS_{1/2}+nP_{1/2}$,
whereas the ground state denoted by $|g\ra$ correlates to  
$nS_{1/2}+nS_{1/2}$ and the van der Waals potential in this latter
case is neglected. Since the atoms are weakly bound by $U(R)$ in
$|e\ra$ we call them {\em quasimolecules}, extrapolating this
denomination to $|g\ra$ as well. The formation of this quasimolecule 
occurs at the Condon point $R_C$, the distance at which $\om_R$ becomes
resonant with an excitation laser of frequency $\om_L<\om_A$. The nuclei 
may then attract each other starting from $R_C$. An eventual spontaneous 
decay from $|e\ra$ to $|g\ra$ releases a
kinetic energy $\Delta K=\hbar\om_L-\hbar\om_\gamma$  
(see Fig~\ref{fig:potential}). If $\Delta K$ is more than twice the
trap depth $V_0$, i.e., if the decay occurs below $R_e$, the pair is
ejected from the trap in the so-called radiative escape collision
\cite{gp89,weiner99}. This two-level approximation neglects hyperfine
splittings of the atomic fine structure and possible (anti) crossings
of potential curves, whereas the nuclear motion is treated semiclassically.
Nevertheless, this turns out to be reliable for a
satisfactory description of cold collisions in the case 
of $^{85}$Rb, a range of hundreds of MHz for the detuning
$\delta=\om_L-\om_A$, and a weak excitation laser intensity 
\cite{gp89,weiner99,peters94,suominen98}. Trapping the atoms in their
ground state, e.g. with a far off-resonance trap (FORT) \cite{fort},
is most suited since optical pumping effects by the trapping light can be
avoided; ejected atoms can then be controlled more efficiently by a
separate excitation light field.

For a high-$Q$ cavity, the main difference with our previous work
\cite{lowq00} is the exchange of excitations between the cavity mode and
the quasimolecules. A set of $N$ identical quasimolecules is 
described by 
\begin{equation}
    H_m=\frac{\hbar\om_R}{2}\sum_{i=1}^N (\sigma_{zi}+1)\, ,
 \label{eq:Hm}
\end{equation}
where each $\sigma_{zi}$ is a Pauli spin matrix. Due to the nuclear
motion, $\om_R$ becomes formally time dependent since $R$ may change 
with time. The actual value of $N$ depends on the detuning $\delta$,
the total number $N_A$ of atoms and their density $n_A$. The
hamiltonian of the quantized field of the cavity mode  
is in turn given by
\begin{equation}
    H_c=\hbar\om_c\, a^\dagger a\, ,
 \label{eq:Hc}
\end{equation}
where $\om_c$ is the cavity resonance frequency and $a^\dagger$ ($a$), the
bosonic creation (annihilation) operator of the field. The interaction
hamiltonian in the rotating-wave approximation is 
\begin{equation}
    H_{int}=\sum_{i=1}^N \left(\frac{\hbar\Omega_i}{2}\,a^\dagger\sigma_i
              + \frac{\hbar\Omega_i^\ast}{2}\,a\,\sigma_i^\dagger\right)\, .
 \label{eq:Hint}
\end{equation}
Here, the Pauli matrices $\sigma_i^\dagger$ and $\sigma_i$ are raising and
lowering operators, respectively, acting  on $|e\ra$ and $|g\ra$ of the i-th
quasimolecule. The individual Rabi frequencies 
\begin{equation}
    \Omega_i=2{\mathcal E}(\om_c)\,f_c({\bf
       r}_i)\,\mbox{\boldmath $\epsilon$}\cdot{\bf d}_i/\hbar
 \label{eq:omega-i}
\end{equation}
depend on the field strength per photon 
${\mathcal E}(\om)=(2\pi\hbar\om/V)^{1/2}$ 
($V$ being the mode volume), polarization {\boldmath $\epsilon$}, mode
profile $f_c({\bf r})$, and the molecular dipole moment ${\bf d}_i$ of the
transition $|g\ra\rightarrow|e\ra$ (whose absolute value $|{\bf
  d}_i|=\sqrt{2}|{\bf d}_A|$, ${\bf d}_A$ being the atomic dipole moment
of $nS_{1/2}\rightarrow nP_{1/2}$).

If the excitation 
laser beam is normal to the cavity axis, the quasimolecules are
excited independently from each other since the wavelength $2\pi
c/\om_R$ is in the optical domain and is much shorter than the average
separation between quasimolecules~\cite{separation}. This corresponds, 
for each quasimolecule, to a no-cavity condition since, in the optical 
domain and for alkalis, typically $\Omega_i \ll \Gamma$, where $\Gamma$ 
is the spontaneous decay 
rate of state $|e\ra$~\cite{haroche82}. In contrast, using for the 
excitation the quantized mode
of the cavity by injecting the laser {\em through} the cavity axis (with
$\om_L=\om_c$), the quasimolecules  
become indistinguishable to the mode $a^\dagger a$ and all of
them have a quantum probability amplitude to be excited. They
therefore end up in a multiparticle entangled excited state. 
Considering only one excitation, it follows from $H_{int}$ that this
state is $|E,0\ra=\sum_i\Omega_i|i,0\ra/\tilde\Omega$,
where $|1,0\ra\equiv|eg\cdots g\ra|0\ra,\ |2,0\ra\equiv|ge\cdots g,\ra|0\ra$
and so forth, with energy $E_e(R)=\hbar\om_R$.  
It couples to the ground state $|G,1\ra \equiv |g\cdots g\ra|1\ra$,
with energy $E_0=\hbar\om_c$ of one field excitation in the mode and all
quasimolecules in their ground state $|g\ra$, by a coupling constant 
\begin{equation}
    \hbar\tilde\Omega=(\sum_i|\hbar\Omega_i|^2)^{1/2}\, .
 \label{eq:omegao}
\end{equation} 
These two states form a closed subspace for the total hamiltonian 
\begin{equation} 
    H=H_m+H_c+H_{int}\, . 
 \label{eq:H} 
\end{equation}

It may be asked whether the evolution of the states $|E,0\ra$ and $|G,1\ra$
under this hamiltonian starts immediately after one excitation is injected into 
the cavity mode. Consider the case of one single pair of free atoms approaching 
each other in the state $|g\ra$ (asymptotic to $S_{1/2}+S_{1/2}$). In the
presence of a light field, the pair may adiabatically pass from $|g\ra$ to
$|e\ra$. If a Landau-Zener approximation \cite{zener32,landau80} holds, the
probability for this to occur is $1-e^{-2\pi\Delta}$, where
$\Delta=\hbar\Omega^2/v_\infty|U^\prime(R_C)|$, $\Omega$ being the coupling
(Rabi frequency) between $|g\ra$ and $|e\ra$, and $v_\infty$, the asymptotic
relative velocity. This approximation is satisfactory for detunings
$|\delta|\gtrsim 10\,\Gamma_A$ and small intensity $\Omega^2$ 
\cite{suominen98}. Analogously, the probability $P_E$ to excite the 
collective state $|E,0\ra$ from $|G,1\ra$ may be estimated by 
\begin{equation}
  \label{eq:exc}
   P_E=1-e^{-2\pi\tilde\Delta}, \hspace{2em}
        \tilde\Delta=\frac{\hbar\tilde\Omega^2}{v_\infty|U^\prime(R_C)|}\, ,
\end{equation}
where the collective coupling $\tilde\Omega$ replaces $\Omega$. Since
$\tilde\Omega>\Omega$, an immediate consequence of the cavity mode is to
increase the excitation probability. Once $|E,0\ra$ is formed, the 
quasimolecules' evolution is then governed by~Eq.(\ref{eq:H}).
Actually, we use a large value for
$\tilde\Omega$ and the Landau-Zener approximation may fail
\cite{suominen98}. Nevertheless, with equal fractions of excited quasimolecules
in both situations with and without cavity, the excitation probability can be  
factored out. Therefore, only the trap loss probability that an {\em excited}
quasimolecule, either in the collective state or in $|e\ra$, is ejected from
the trap becomes significant.

The most notable feature arises as time goes on. One expects a Rabi
oscillation between $|E,0\ra$ and $|G,1\ra$ as the excitation is
exchanged between the quasimolecules and the cavity mode. This implies
that the quasimolecules switch between $|e\ra$ and $|g\ra$ at the
collective rate $\tilde\Omega$, altering completely the collisional
encounter compared to how it would proceed without the quantum field; 
and the larger the number of quasimolecules $N$, the
faster such oscillation will be , since
$\tilde\Omega$ scales with $\sqrt{N}$. This oscillation looses
strength as $\om_R$ detunes gradually from $\om_c$. When
$|\om_R-\om_c|\gtrsim\tilde\Omega$, we may neglect $H_{int}$ and let
the quasimolecules evolve freely, subject at most to spontaneous decay. 
It is interesting to mention here that the resonance condition is 
determined by $\tilde\Omega$ instead of the cavity (or molecular) 
linewidth. This is analogous to a power broadening effect, even though 
we consider a single excitation in the cavity: the ``power'' here is 
increased via an increase of the collective dipole. 
If there is a probability that the quasimolecules remain excited on entering
the off-resonance region $R<R^\prime$, determined by
\begin{equation}
  \label{eq:R-prime}
   \om_{R_C}-\om_{R^\prime}\approx\tilde\Omega\, ,
\end{equation}
they may reach $R<R_e$ where spontaneous decay of $|E,0\ra$ ejects a pair
from the trap. In contrast, if the interaction time is such that the
quasimolecules leave the resonant region $R^\prime<R<R_C$ in the ground
state $|G,1\ra$ (having thus undergone a so-called $\pi$-pulse),
practically no trap loss will be recorded at all. From the dependence of this
interaction time on the detuning $\delta$, we deduce in the following an
expression for the trap-loss probability as a means to observe and measure
this collective Rabi oscillation of colliding cold atoms as a function of 
$\delta$.

If $t_c \lesssim 2\pi/\tilde\Omega$ is the interaction time spent in the
resonance region $R^\prime<R<R_C$, we keep the resonance condition 
$\om_R\approx \om_c$, since
$|\om_R-\om_c|t_c \lesssim \tilde\Omega\times 2\pi/\tilde\Omega\sim 2\pi$ 
completes at most a single cycle. Analogously, the time variation of
$\dot{\om}_R\neq 0$ is neglected: the phase $\theta(t_c)\equiv
\int_0^{t_c} d\tau\,|\dot{\om_R}|\tau$ that such time dependence would
introduce to the dynamics is at most of order $2\pi$ if we use that
$t_c \lesssim 2\pi/\tilde\Omega$ and $|\Delta\om_R| \lesssim
\tilde\Omega$. A full solution is possible, but this suffices for our
purposes, limiting us, however, to few oscillations. The
spontaneous decay rate of the state $|e\ra$ is $\Gamma=2\Gamma_A$,
where $1/\Gamma_A$ is the lifetime of the atomic state $nP_{1/2}$
\cite{power67}, and retardation effects can be neglected ($R$ is
safely smaller than $2\pi c/\om_R$). Since the cavity mode solid angle
is small, the rate of emission into the rest of free space is nearly
$\Gamma$. For the quasimolecules and the cavity mode, the Liouville-von
Neumann equation with dissipation is 
\begin{equation}
     \dot{\rho} = \frac{1}{i\hbar}[H,\rho] + \sum_{i=1}^N \frac{\Gamma}{2}
     (2\sigma_i\rho\sigma_i^\dagger - \sigma_i^\dagger\sigma_i\rho
           - \rho\sigma_i^\dagger\sigma_i)\, ,
 \label{eq:liouville}
\end{equation}
where the cavity dissipation is neglected. The dissipator in this
equation assumes an incoherent decay of each quasimolecule
independently from each other, since the emitted wavelength is shorter
than the average separation of quasimolecules~\cite{separation}. When
the emission is towards the cavity mode, in contrast, a collective
effect in the quasimolecules' emission is fully contained in the
interaction with the field.

For the initial condition $\rho(0)=|E,0\ra\la E,0|$, we calculate the
probability  
\begin{equation}
    p_E(t)={\rm tr}\left[|E,0\ra\la E,0|\rho(t)\right]
 \label{eq:p}
\end{equation}
that the system remains in the state $|E,0\ra$ after a time interval $t$.
The solution can be found by enlarging the subspace spanned by
$|E,0\ra$ and $|G,1\ra$ to include the vacuum state 
$|V\ra\equiv|G,0\ra\equiv |gg\cdots g\ra|0\ra$ to which the system evolves 
due to dissipation. In the
interaction representation, one has
\begin{eqnarray}
  \label{eq:rho}
  \rho_I(t) &=& e^{i(H_m+H_c)t/\hbar}\rho(t)e^{-i(H_m+H_c)t/\hbar}
                                                  \nonumber \\
            &=& p_E |E,0\ra\la E,0| + p_G |G,1\ra\la G,1| 
                          + p_V |G,0\ra\la G,0|  \nonumber \\
            & &\hspace{1.0cm}  +  \left(  c_{EG} |E,0\ra\la G,1|
                                + c_{EV} |E,0\ra\la G,0| \right.
                                                  \nonumber \\
        & &\hspace{2.5cm}\left. + c_{GV} |G,1\ra\la G,0| + {\rm adj.}
                           \right)\, .
\end{eqnarray}
The time-dependent probabilities $p_E$, $p_G$, and $p_V$, and the
coherences $c_{EG}$, $c_{EV}$, and $c_{GV}$ are found by substitution
of $\rho_I(t)$ into the interaction representation of the
Liouville-von Neumann equation and projecting into each component of 
$\rho_I(t)$. The damping terms generate no new state and we get for
the $p$'s 
\begin{eqnarray}
  \dot{p}_E &=& - \Gamma\,p_E + i\tilde\Omega\,(c_{EG} -  c^\ast_{EG}),
                                                         \label{eq:pe}\\   
  \dot{p}_G &=& - i\tilde\Omega\,(c_{EG} - c^\ast_{EG}),
                                                         \label{eq:pg}\\
  \dot{p}_V &=& \Gamma\,p_E\, ,                             \label{eq:p0}
\end{eqnarray}
and for the $c$'s
\begin{eqnarray}
  \label{eq:cs}
  \dot{c}_{EG} &=& - \frac{\Gamma}{2}\,c_{EG} + i\tilde\Omega\,(p_E - p_G),
                                                         \label{eq:ceg}\\ 
  \dot{c}_{EV} &=& - \frac{\Gamma}{2}\,c_{EV} - i\tilde\Omega\,c_{GV},
                                                         \label{eq:ce0}\\
  \dot{c}_{GV} &=& - i\tilde\Omega\,c_{EV}\, .              \label{eq:cg0}
\end{eqnarray}
From the last two equations, it follows that $c_{EV}(t)=c_{GV}(t)=0$
if this holds initially. From Eq.(\ref{eq:pe}), Eq.(\ref{eq:pg}), and
Eq.(\ref{eq:ceg}), $p_E(t)$ can be solved straightforwardly.
Since $\tilde\Omega>\Gamma/4$ (condition for Rabi oscillations) is 
supposed to be fulfilled with the parameters we adopt, the solution is 
the {\em underdamped} one
\begin{equation}
  \label{eq:+}
  p_\Omega(t)\equiv p_E(t)=e^{-\Gamma t/2}
      \left(\cos{\beta t} 
           - \frac{\Gamma}{4\beta}\sin{\beta t}\right)^2\, ,
\end{equation}
where 
\begin{equation}
  \label{eq:beta}
  \beta\equiv \sqrt{\tilde\Omega^2 - (\Gamma/4)^2}, 
                          \hspace{2em} \tilde\Omega>\Gamma/4\, .
\end{equation}
As already mentioned, this solution is to be applied so long as the
resonance condition $|\om_R-\om_c|\lesssim\tilde\Omega$ holds, that
is, as the quasimolecules-field interaction is significant. For 
$|\om_R-\om_c|\gtrsim\tilde\Omega$, on the other hand, we neglect this
interaction. The solution then is {\em formally} equivalent to taking
the limit $\tilde\Omega\longrightarrow 0$ in the {\em overdamped}
regime, which follows from Eq.(\ref{eq:+}) by substituting
$-i|\beta|$ for $\beta$ (i.e., solving Eq.(\ref{eq:liouville}) without
$H_{int}$), 
\begin{equation}
  \label{eq:pover}
  p_E(t)\approx e^{-\Gamma t}, \hspace{2em} 
                     |\om_R-\om_c|\gtrsim\tilde\Omega\, .
\end{equation}
As expected, this equation implies the value $\Gamma$ for the 
decay rate of the collective state $|E,0\ra$, the same as the 
molecular value for $|e\ra$. This can be derived directly from a
perturbation calculation {\em \`a la} Fermi's golden rule by coupling
the quasimolecules to all the electromagnetic vacuum modes except the
cavity mode $\om_c$, whose small solid-angle is neglected. It reflects
the fact that the average separation between quasimolecules is larger
than the wavelength of the emitted radiation, and the indistinguishability of
which quasimolecule emits into free space can no longer
hold~\cite{separation}.

We need $p_\Omega$ to obtain the probability that one pair of atoms is ejected 
from the trap after excitation by the cavity mode. The transition from
$|E,0\ra$ to $|G,1\ra$ should then occur for $R<R_e$. In the first passage in
this region, the probability $l_1$ that a pair of atoms from {\em any} one of 
the $N$ quasimolecules is ejected is constructed as
\begin{eqnarray}
  \label{eq:L1}
    l_1&=&\sum_{i=1}^N \left|\frac{\Omega_i}{\tilde\Omega}\right|^2
               p_\Omega(t_c)\,e^{-t^\prime\Gamma}\,(1-e^{-2t_e\Gamma})
                                                       \nonumber \\
             &=& p_\Omega(t_c)\,e^{-t^\prime\Gamma}\,(1-e^{-2t_e\Gamma})\, ,
\end{eqnarray}
where $t_e$ ($t^\prime$, or $t_c$) is the time interval spent between 
$R=0$ and $R_e$ ($R_e$ and $R^\prime$, or $R^\prime$ and $R_C$). This 
$l_1$ is thus a sum of products of conditional probabilities \cite{gp89}, 
which in our 
case involve Eq.(\ref{eq:+}) and Eq.(\ref{eq:pover}). Now, by letting $R$
evolve further, the quasimolecules can vibrate between $R_C$ and $R=0$ before 
emission takes place, that is, they may pass several times across
$R_e$ while still in $|E,0\ra$. Summing over these multiple passages
\cite{peters94} and composing them with conditional probabilities, the
probability ${\cal L}_c$ for a pair of atoms to be ejected at any time
is then 
\begin{eqnarray}
  \label{eq:Lc}
   {\cal L}_c &=&  l_1 + l_2 + l_3 + \cdots                \nonumber\\
   &=&\sum_{i=1}^N \left|\frac{\Omega_i}{\tilde\Omega}\right|^2
            \left\{p_\Omega(t_c)\,e^{-t^\prime\Gamma}\,(1-e^{-2t_e\Gamma})
                                                    \right.\nonumber\\
   & &\mbox{} + p_\Omega(t_c)\,
                \left[e^{-2(t^\prime + t_e)\Gamma}\,p_\Omega(2t_c)\right]
                \,e^{-t^\prime\Gamma}\,(1-e^{-2t_e\Gamma}) \nonumber\\
   & &\mbox{} +  p_\Omega(t_c)\,
                \left[e^{-2(t^\prime + t_e)\Gamma}\,p_\Omega(2t_c)\right]^2
                        \,e^{-t^\prime\Gamma}\,(1-e^{-2t_e\Gamma}) \nonumber\\
   & &\mbox{} + \cdots \left. \right\}                    \nonumber\\
   &=& p_\Omega(t_c)\,\frac{\sinh{(\Gamma t_e)}}{\frac{1}{2}
                \left[e^{(t^\prime + t_e)\Gamma} - 
                 p_\Omega(2t_c)\,e^{-(t^\prime + t_e)\Gamma}\right]}\, .
\end{eqnarray}
As in Eq.(\ref{eq:L1}), the summation over the $N$ quasimolecules is
factored since the multiple passages probabilities $l_n$ are the same for
each quasimolecule. In $l_n$, $p_\Omega(t_c)$ is the probability that a
quasimolecule exits the region $R^\prime<R<R_C$ in the excited state $|e\ra$
after having been excited at $R_C$, whereas $p_\Omega(2t_c)$ gives the
probability of crossing $R^\prime$ in $|e\ra$ from the right after having
crossed it from the left in $|e\ra$.

The dependence of ${\cal L}_c$ on the collective Rabi frequency
$\tilde\Omega$ can be observed formally by changing the
interaction time $t_c$ and thus making $p_\Omega$ oscillate. This time
interval in turn is a function of the detuning $\delta=\om_c-\om_A$
and can be obtained by integration of the energy conservation
condition \cite{gp89}   
\begin{equation}
  \label{eq:energy}
  \frac{\mu\dot{R}^2}{2} + U(R) = {\rm const.}\, .
\end{equation}
Indeed, choosing $\tilde\Omega$ such that $t^\prime\approx 0$ (i.e.,
$R^\prime\approx R_e$, see Eq.(\ref{eq:R-prime})), we have 
$\hbar\tilde\Omega = 2V_0$ (emission on resonance with the cavity 
will not lead to trap-loss) and neglecting the
initial velocity $\dot{R}$ at $R=R_C$, it follows from Eq.(\ref{eq:energy}) 
\begin{eqnarray}
  \label{eq:time}
  t_c &=& t_0(\delta)\,f(\delta), \hspace{2em} 
                          (t^\prime\approx 0, R^\prime\approx R_e)
                                                 \nonumber \\
  t_e &=& t_0(\delta)\,[1-f(\delta)]\, ,
\end{eqnarray}
where $t_0=t_c+t_e$ is the total time interval between $R_C$ and
$R=0$, and $f(\delta)$ is the fraction of $t_0$ spent between $R_C$
and $R_e$, 
\begin{eqnarray}
   t_0(\delta) &=& g_0\left(\frac{\mu}{2C_3}\right)^{1/2} 
             \left(\frac{C_3}{\hbar|\delta|}\right)^{5/6},       \\
   f(\delta) &=& \frac{1}{g_0}\int_r^1du\frac{1}{\sqrt{u^{-3}-1}},
   \hspace{0.2em} r\equiv \frac{R_e}{R_C} =
                 \left(1+\frac{\tilde\Omega}{|\delta|}\right)^{-1/3},
                                                         \nonumber
\end{eqnarray}
$g_0=0.746$ normalizing $f(0)$ to unity. A slight complication is
the dependence of $\tilde\Omega$ on $\delta$ as well via the total
number $N=N(\delta)$ of quasimolecules, so that $\tilde\Omega$ changes
as $t_c(\delta)$ is changed. In order to obtain
$\tilde\Omega=\tilde\Omega(\delta)$, we approximate for large $N$ the summation 
in Eq.(\ref{eq:omegao}) by an average (${\bf e}_i$ being the orientation of
${\bf d}_i$, and $d$ its modulus)
\begin{eqnarray}
  \label{eq:od}
   \tilde\Omega^2 &=& \sum_i |\Omega_i|^2                 \nonumber \\
        &=& N \frac{{\cal E}^2 d^2}{\hbar^2}\left(\frac{1}{N}
        \sum_i|f_c({\bf r}_i)|^2 |\mbox{\boldmath $\epsilon$}
                                \cdot {\bf e}_i|^2\right)\nonumber \\
     &\approx&  N \frac{{\cal E}^2 d^2}{\hbar^2}\la|f_c({\bf r}_i)|^2\ra
        \la|\mbox{\boldmath $\epsilon$}\cdot {\bf e}_i|^2\ra \nonumber \\
     &\sim&  N \frac{{\cal E}^2 d^2}{6\hbar^2} \equiv N\Omega^2\, ,
\end{eqnarray}
with $|f_c({\bf r}_i)|\sim\cos{(z_i k_c)}$ ($z$ along the cavity axis and
\hbox{$k_c=\om_c/c$}) and ${\bf e}_i$ being randomly oriented, and $\Omega$
denoting an (averaged) single quasimolecule Rabi frequency. The total number
$N$ can be estimated by counting all pairs of atoms whose separation $R$ is
such that $\om_R=\om_c$, with a spread $\Delta R$ about the Condon Point
$R_C$ determined by the linewidth $\Gamma$ of the state $|e\ra$, namely, 
\begin{eqnarray}
  \label{eq:N}
   N &\sim& \frac{1}{2}N_A n_A\,4\pi R_C^2 \Delta R  \nonumber \\
     &\sim& \frac{1}{2}N_A n_A\,4\pi R_C^2 
                \frac{\Gamma}{|d\om_R/dR|_{R_C}} \nonumber \\
     &\sim& N_A n_A\,\frac{2\pi C_3}{3\hbar\Gamma}\,
                \left(\frac{\Gamma}{\delta}\right)^2\, , 
\end{eqnarray}
having used that $C_3/R_C^3=\hbar|\delta|$.

For the sake of comparison, we calculate the trap loss probability
$\cal{L}_{\circ}$ that, in the {\it absence} of the cavity, $|e\ra$ 
decays in the region
$R<R_e$ after multiple passages across $R_e$. Both ${\cal L}_c$ and 
${\cal L}_{\circ}$ are similar, except by $p_\Omega(t)$ which is substituted 
for the pure decay
$p_E(t)\approx e^{-t\Gamma}$   
\begin{equation}
  \label{eq:lo}
  {\cal L}_o = \frac{\sinh{(t_e\Gamma)}}{\sinh{(t_c+t_e)\Gamma}}\, ,
\end{equation}
describing the trap loss probability of a statistical mixture of 
pairs of atoms colliding independently~\cite{weiner99,peters94}.   

\begin{figure}
\centering \resizebox{8.0cm}{!}{\includegraphics*{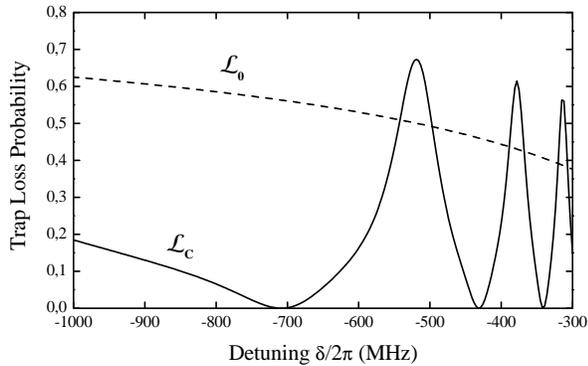}}
\vspace{1em}
\caption{Trap loss probabilities ${\cal L}_c$ and $\cal{L}_{\circ}$ as
functions of the detuning $\delta=\om_c-\om_A$ (see text for details). Note the 
``oscillation'' in ${\cal L}_c$ brought about by the interaction with the 
cavity quantized mode. The curve levels off and approaches $\cal{L}_{\circ}$
for $\delta/2\pi<-800$~MHz, where the interaction time $t_c$ goes to zero.}
\label{fig:loss}
\end{figure}

For numerical estimates, we consider $^{85}$Rb atoms and the atomic
transition $5S_{1/2}\rightarrow5P_{1/2}$. The wavelength of the transition is 
$\lambda_A=2\pi c/\om_A=795$~nm, with atomic decay rate $\Gamma_A/2\pi =
6$~MHz. The coefficient of the dipole-dipole potential is taken as
$C_3\approx 11\times 10^{-11}$~erg$\,$\AA$^3$ \cite{bussery85,cline94}
and the detunning $\delta=\om_c-\om_A<0$ is made to decrease, starting
from $\delta/2\pi=-350$~MHz. For this $\delta$, the Condon point is
$R_C=366$~\AA\ and the total time $t_0= 1.07\times 10^{-8}$~s. The
trap depth $V_0$ is chosen of order $5$~mK ($\sim100$~MHz) and would have to
decrease as $\delta$ is decreased in order to match the condition (see
Eq.(\ref{eq:energy}) {\em et seq.}) 
\begin{equation}
  \label{eq:omega}
     2V_0=\hbar\tilde\Omega, \ \ {\rm or}
        \ \ \tilde\Omega/2\pi\lesssim 200\ {\rm MHz}\, .
\end{equation}
This choice for $\tilde\Omega$ allows for the quasimolecules to enter
directly the trap escape region $R<R_e$ as they exit the resonant
region $R_e<R<R_C$ in which their interaction with the cavity mode is
strongest. With such parameters, and $\delta$ going down to $\sim
-1.0$~GHz (when $\tilde\Omega/2\pi\approx 70$~MHz), we can approximate
$\beta\approx \tilde\Omega$ and neglect the sine in Eq.(\ref{eq:+}),

\begin{equation}
  p_\Omega(t_c)\approx e^{-t_c\Gamma/2}\cos^2{(\tilde\Omega t_c)}.
\end{equation}

\noindent
For the detuning $\delta=-350$~MHz, one obtains $t_c=t_0 f = 0.55t_0$ 
and $t_e=(1-f)t_0=0.45t_0$. The largest phase becomes then
$\tilde\Omega t_c\approx 
2\pi\times 200{\rm MHz}\times 0.55\times 1.07\times 10^{-8}{\rm
  s}=2.35\pi$ and $\tilde\Omega\,t_c$ will decrease as $\delta$ decreases.
In Fig.(\ref{fig:loss}), we plot $\cal{L}_{\circ}$ and ${\cal L}_c$, assuming
the quasimolecules are equally excited in both cases so that the
excitation probabilities can be factored out. Outside the range of
detunings shown, the approximations in our model are no longer valid 
\cite{weiner99,suominen98}: for smaller $|\delta|$, a full quantum
mechanical treatement of the nuclear dynamics is required, whereas
the discrete quantum vibrational levels of the dipole-dipole potential
$U(R)$ become resolved for $|\delta|\gtrsim 1.0$ GHz.

Within this detuning window, nevertheless, it can be clearly seen the
drastic change brought about by the cavity mode. The collective Rabi
oscillations begin mildly at a large $|\delta|$ and become more pronounced
at smaller values of absolute detuning, where the interaction time $t_c$ is 
longer. At the minima of ${\cal L}_c$, the quasimolecules leave the resonant
region in the state $|G,1\ra$ (having undergone a $\pi$-pulse) and the 
kinetic energy is insufficient to eject a pair of atoms from the trap.
The maxima of ${\cal L}_c$, where the quasimolecules leave the resonant
region in the excited state, are above $\cal{L}_{\circ}$ since, without the
cavity effect, the probability of a quasimolecule to reach the escape region in 
the excited state $|e\ra$ is smaller. These maxima increase with decreasing 
$\delta$ because the potential becomes steeper and, therefore, the time 
$t_c$ decreases. The effect of damping during the interaction with the 
cavity mode becomes less and less important and $p_{\Omega}$ becomes closer 
to one (see Eqs.(\ref{eq:+}) and (\ref{eq:Lc})).

It should be noted that an actual realization of such an experiment is quite
difficult. For a centimeter sized optical resonator with mirrors separated by 
$l\approx 1$~cm, the mode volume $V\approx\pi w_o^2l\approx 4.0\times
10^{-5}$~cm$^3$, with a waist $w_o=\sqrt{c l/\om_c}\approx 36\,\mu$m, so 
that using $d=\sqrt{2}\,d_A$ and $\Gamma_A=4d_A^2\om_A^3/3\hbar c^3$
($\om_A\approx \om_c$), the averaged single quasimolecule Rabi frequency in
Eq.(\ref{eq:od}) is $\Omega/2\pi\approx 0.42$~MHz. From Eq.(\ref{eq:omega})
and Eq.(\ref{eq:od}), one needs at least $N\sim (\tilde\Omega/\Omega)^2\approx
2.3\times 10^5$ quasimolecules (but less for larger $|\delta|$). Using
Eq.(\ref{eq:N}), this could only be achieved with a gas of $N_A\sim 2.0\times
10^9$ atoms occupying all the mode volume $V$, i.e., at a high density
$n_A\sim 4.0\times 10^{13}$~cm$^{-3}$. A possible remedy that could do with
less atoms at lower densities would imply, however, a smaller $\tilde\Omega$ 
and thus a substantially longer
interaction time $t_c$ in order for the phase $\tilde\Omega t_c$ to reach
$\sim 2\pi$ and show a ``collisional Rabi oscillation''; for these longer
times, the detuning would fall below the limit $|\delta|\gtrsim
10\,\Gamma_A$ and our semiclassical approximation for the nuclear motion
would break down~\cite{suominen98}.

In conclusion, it is a remarkable effect that the outcome of a collision
encounter between two cold atoms can be so deeply affected when it takes place
in the presence of a quantized cavity mode. The slowness of the collision
makes possible the exchange of excitation between the colliding pair of
atoms and the cavity field. Depending on how long the interaction is
effective, the potential energy of the collision can simply be ``turned
off'' as the quasimolecule leaves the excitation within the cavity; in
contrast, the quasimolecule may be taken closer to the radiative escape
condition if this same excitation is returned back to it. In this process,
several quasimolecules get quantum mechanically entangled to each other. 
The trap-loss probability as a function of the detuning can thus show a
Rabi-like collective oscillation.

We acknowledge financial support from the Brazilian agencies FAPESP and  
CNPq and thank R. B. B. Santos for stimulating discussions.





\begin{thebibliography}{99}

\bibitem{yoo85} H.-I. Yoo, J.H. Eberly, Phys. Rep. {\bf 118} (1985) 239.

\bibitem{jc63} E.T. Jaynes, F.W. Cummings, Proc. IEEE {\bf 51} (1963) 89.

\bibitem{haroche96} M. Brune, F. Schmidt-Kaler, A. Maali, J. Dreyer, E. 
Hagley, J.M. Raimond, S. Haroche, Phys. Rev. Lett. {\bf 76} (1996) 1800. 

\bibitem{wine96} D.M. Meekhof, C. Monroe, B.E. King, W.M. Itano, 
D.J. Wineland, Phys. Rev. Lett. {\bf 76} (1996) 1796.

\bibitem{dicke54} R.M. Dicke, Phys. Rev. {\bf 93} (1954) 99.

\bibitem{bonifacio71} R. Bonifacio, P. Schwendimann, F. Haake,
  Phys. Rev. {\bf A4} (1971) 302; {\em ibid.} {\bf A} (1971) 854.

\bibitem{gibbs81} M.F.H. Schuurmanns, Q.H.F. Vrehen, D. Polder,
  H.M. Gibbs, Adv. At. Mol. Phys. {\bf 17} (1981) 167.

\bibitem{gross82} M. Gross, S. Haroche, Phys. Rep. {\bf 93} (1982) 301.

\bibitem{haroche82} S. Haroche, in: G. Grynberg, R. Stora (Eds.), Les Houches,
  Session XXXVIII, Vol. I, North Holland, Amsterdam, 1984.

\bibitem{gp89} A. Gallagher, D.E. Pritchard, Phys. Rev. Lett. 
  {\bf 63} (1989) 957.

\bibitem{weiner99} J. Weiner, V.S. Bagnato, S. Zilio, 
  P.S. Julienne, Rev. Mod. Phys. {\bf 71} (1999) 1.

\bibitem{lowq00} J.I. Kim, R.B.B. Santos,  P. Nussenzveig,
  Phys. Rev. Lett. {\bf 86} (2001) 1474.

\bibitem{mossberg91} T.W. Mossberg, M. Lewenstein,  D.J. Gauthier,
  Phys. Rev. Lett. {\bf 67} (1991) 1723.

\bibitem{gangl00} M. Gangl, H. Ritsch, Phys. Rev. {\bf A61} (2000) 043405.

\bibitem{deb99} B. Deb, G. Kurizki, Phys. Rev. Lett. {\bf 83} (1999) 714.  

\bibitem{peters94} M.G. Peters, D. Hoffmann, J.D. Tobiason, 
  T. \hbox{Walker}, Phys. Rev. {\bf A50} (1994) R906.

\bibitem{suominen98} K.-A. Suominen, Y.B. Band, I. Tuvi, K. Burnett,
 P.S. Julienne, Phys. Rev. {\bf A57} (1998) 3724.

\bibitem{fort} S. Chu, J. Bjorkholm, A. Ashkin,  A. Cable,
  Phys. Rev. Lett. {\bf 57} (1986) 314; J.D. Miller, R.A. Cline, 
  D. J. Heinzen, Phys. Rev. A {\bf 47} (1993) R4567. 

\bibitem{separation} From the density $n\approx N/V\approx 5.7\times
  10^9$~cm$^{-3}$ of quasimolecules estimated at the end of this
  paper, one gets $n^{-1/3}\approx 5600$~nm for the average 
  separation between quasimolecules.

\bibitem{zener32} C. Zener, Proc. R. Soc. {\bf A137} (1932) 696.

\bibitem{landau80} L. Landau, E. Lifchitz, {\em M\'ecanique Quantique},
3$^{\mbox{\scriptsize \`eme}}$ \'edition, \'Editions Mir, Moscou, 1980.

\bibitem{power67} E.A. Power, J. Chem. Phys. {\bf 46} (1967) 4297.
 
\bibitem{bussery85} B. Bussery, M. Aubert-Fr\'econ,
  J. Chem. Phys. {\bf 82} (1985) 3224.

\bibitem{cline94} R.A. Cline, J.D. Miller,  D.J. Heinzen,
  Phys. Rev. Lett. {\bf 73} (1994) 632.


\end{thebibliography}
\end{document}